# Title

**Segmentation of spinal rootlets across MRI contrasts with RootletSeg**

# Abbreviation key

PMJ = pontomedullary junction, SD = standard deviation, T1w = T1-weighted, T2w = T2-weighted

# Abstract

Segmentation of spinal nerve rootlets is relevant for spinal level estimation, lesion classification, neuromodulation therapy, and group-level analyses. The aim of this study was to develop a deep learning method for the automatic segmentation of C2-T1 dorsal and ventral spinal nerve rootlets on various MRI scans. The study included MRI scans from two open-access and one private dataset, consisting of 3D isotropic 3T turbo spin echo T2-weighted (T2w) and 7T MP2RAGE (T1-weighted [T1w] INV1 and INV2, and UNIT1) MRI scans. A deep learning model, *RootletSeg*, was developed on 93 MRI scans from 50 healthy adults (mean age, 28.70 years ± 6.53 [SD]; 28 [56%] males, 22 [44%] females) and achieved a mean ± SD Dice score of 0.67 ± 0.09 for T1w-INV2, 0.65 ± 0.11 for UNIT1, 0.64 ± 0.08 for T2w, and 0.62 ± 0.10 for T1w-INV1 contrasts. RootletSeg accurately segmented C2-T1 spinal rootlets across MRI contrasts, enabling the determination of spinal levels directly from MRI scans. The method is open-source and can be used for a variety of downstream analyses.

# Title Page

**List of authors**

*Katerina Krejci[1,2*], Jiri Chmelik[1], Sandrine Bédard[2], Falk Eippert[3], Ulrike Horn[3], Virginie Callot[4,5], Julien Cohen-Adad[2,6,7,8], Jan Valosek[2,9,10,11*]*

* corresponding authors

**Affiliations:**

1. Department of Biomedical Engineering, FEEC, Brno University of Technology, Brno, Czechia
2. NeuroPoly Lab, Institute of Biomedical Engineering, Polytechnique Montreal, Montreal, QC, Canada
3. Max Planck Research Group Pain Perception, Max Planck Institute for Human Cognitive and Brain Sciences, Leipzig, Germany
4. Aix Marseille Univ, CNRS, CRMBM, Marseille, France
5. APHM, CHU Timone, Pôle d'Imagerie Médicale, CEMEREM, Marseille, France
6. Mila - Quebec AI Institute, Montreal, QC, Canada
7. Functional Neuroimaging Unit, CRIUGM, Université de Montréal, Montreal, QC, Canada
8. Centre de Recherche du CHU Sainte-Justine, Université de Montréal, Montreal, QC, Canada
9. Department of Neurosurgery, Faculty of Medicine and Dentistry, Palacký University Olomouc, Olomouc, Czechia
10. Department of Neurology, Faculty of Medicine and Dentistry, Palacký University Olomouc, Olomouc, Czechia
11. Spinal Cord Injury Center, Balgrist University Hospital, University of Zurich, Zurich, Switzerland

**Corresponding author address**

Jan Valosek, PhD, jan.valosek@upol.cz, Palacký University Olomouc, Faculty of Medicine and Dentistry, Hněvotínská 976/3, Nová Ulice, 779 00, Olomouc, Czechia

# Introduction

Spinal rootlets are bundles of nerve fibres forming the spinal nerves that connect the spinal cord to the peripheral parts of the body. The ability to estimate neurological spinal levels from nerve rootlets makes them relevant for spinal cord lesion classification[1,2], neuromodulation therapy[3,4], and fMRI group analysis[5,6,7]. Because directly identifying spinal rootlets on MRI scans is both challenging and time-consuming, spinal cord analyses typically rely on vertebral levels, defined using intervertebral discs, or they infer spinal levels from vertebral levels. This approach, however, is intrinsically limited as spinal levels are not necessarily aligned with vertebral bodies[5,8], and there is considerable inter-individual variability between spinal and vertebral levels[8–13]. Spinal nerve rootlets, therefore, provide an anatomically more relevant and potentially more accurate method for determining spinal levels, which can, in turn, serve as an alternative coordinate system for spinal cord analyses, as opposed to the traditional vertebral-level approach based on intervertebral discs[5]. Recently, a method for automatic spinal rootlets segmentation was proposed, allowing for direct estimation of spinal levels from MRI data[14]. However, that method has three drawbacks: i) it was developed solely on scans acquired at one field strength and with one contrast (i.e. 3T turbo spin echo [TSE] isotropic T2-weighted [T2w] scans), ii) it is restricted to dorsal rootlets only (ignoring ventral rootlets), and iii) it is restricted to a specific range of cervical levels (i.e. spinal levels C2-C8). Other studies aimed to identify nerve rootlets using diffusion MRI tractography[15] or traced them manually on high-resolution scans[16]. An automatic rootlets segmentation method was recently proposed also for postmortem feline samples[17,18].

In this work, we (i) extend the existing rootlet segmentation model by incorporating ventral rootlets, an additional spinal level (thoracic level T1), and additional MRI contrasts derived from the 7T MP2RAGE sequence (T1w-INV1, T1w-INV2, and UNIT1); and (ii) utilize the proposed segmentation model to investigate the correspondence between spinal and vertebral levels in a large cohort of 120 healthy participants. The segmentation method is open-source, implemented in the `sct_deepseg` function as part of Spinal Cord Toolbox (SCT)[19] v7.0 and higher.

# Materials and Methods

## Study Design and Participants

This retrospective study included scans from three MRI datasets of the cervical spinal cord (Suppl. Table 1): (i) 3T TSE isotropic T2w scans from the open-access single-site OpenNeuro ds004507 dataset[20], (ii) 3T TSE isotropic T2w scans from the open-access spine-generic multi-subject dataset[21], and (iii) 7T isotropic MP2RAGE scans from a private single-site dataset[22,23]. For more details on acquisition parameters, please see[20,22,24,25]. The inclusion and exclusion criteria are listed in the Supplementary material (1.1. Study Design and Participants).

The experimental protocols for all datasets were approved by the relevant institutional ethics committees prior to data acquisition, specifically: the Comité d'éthique de la recherche du Regroupement Neuroimagerie Québec (OpenNeuro ds00450720 dataset[20]), local ethics committees of the participating institutions (spine-generic multi-subject dataset[21]) and by the Ethics Committee of the Medical Faculty of the University of Leipzig (private single-site dataset[22]). All datasets complied with all relevant ethical regulations, all experiments were performed in accordance with the Declaration of Helsinki and informed consent was obtained from each participant.

## Deep Learning Training Protocol

The nerve rootlets segmentation model was developed using `nnUNetv2`, a self-configuring deep learning-based framework[26]. Details on data preprocessing and augmentation can be found in the Supplementary material (1.2. Deep Learning Training Protocol).

To facilitate the generation of reference standard rootlet labels, the existing segmentation model was applied[14], followed by manual corrections by consensus of two raters (K.K. and J.V.) using the FSLeyes image viewer (version 1.12.1; University of Oxford) and the provided instructions[1]. We did not analyze the interrater variability, as a previous study reported a mean coefficient of variation of ≤ 1.45% in spinal level positions when using rootlet segmentations from different raters to estimate spinal levels[14]. Additionally, as the presence of C1 rootlets differs between individuals, they were not included in reference standard annotations and model training[12,14,27]. We trained the model in four stages by iteratively adding more MRI scans and contrasts in each stage (Figure 1). Detailed training procedure is described in the Supplementary material (1.2. Deep Learning Training Protocol). During model development, we observed that increasing the training patch size was necessary to successfully segment the C2 level, suggesting that a larger spatial context is important for accurate rootlet segmentation at this level. The final RootletSeg model configuration is included as a JSON file in the GitHub release (https://github.com/ivadomed/model-spinal-rootlets/releases/tag/r20250917). The model's performance was evaluated using the Dice score and the 95th percentile Hausdorff distance (HD95)[28], where a higher Dice score and a lower HD95 indicate better segmentation performance.

[1] https://github.com/ivadomed/model-spinal-rootlets/issues/17

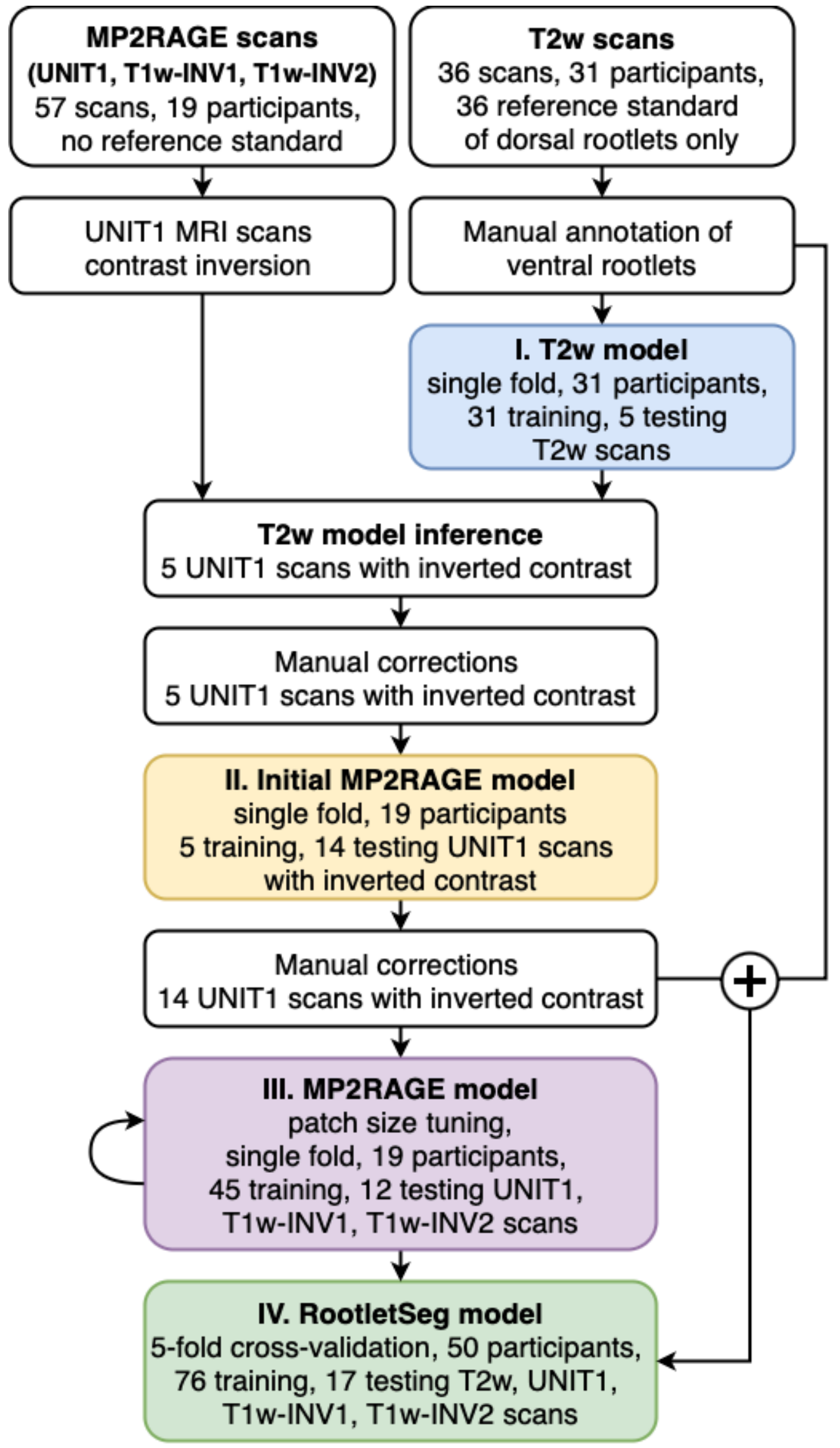

**Figure 1: Overview of our model development.** T2-weighted (T2w) scans with reference standards were first used to train the T2w model (I), which was then applied to MP2RAGE-UNIT1 scans with inverted contrast. The resulting outputs were manually corrected and served as reference standards for the training of the initial MP2RAGE model (II). This model was applied to additional UNIT1 scans, followed by another round of manual corrections. Next, the MP2RAGE model (III), using both the original and an increased patch size, was trained on all three MP2RAGE contrasts (T1w-INV1, T1w-INV2, and UNIT1). Finally, T2w and MP2RAGE scans and their corresponding reference standards were combined into a single dataset and used to train the *RootletSeg* model.

## Spinal-vertebral Level Analysis

The developed model was used to analyze the correspondence between spinal and vertebral levels. For MP2RAGE data, where each participant has three perfectly aligned contrasts, the T1w-INV2 contrast was used to avoid repeating the same participants multiple times. T1w-INV2 was selected due to its superior rootlet visibility and the highest *RootletSeg* performance among the MP2RAGE contrasts (see Results). From the OpenNeuro dataset, only participants with a neutral neck position were included to avoid repeating the same participants across different sessions. From the spine-generic dataset, MRI scans with clear difference in intensity values between the cerebrospinal fluid and nerve rootlets were selected. To be included in the analysis, each scan had to meet the following criteria: good visibility of spinal rootlets (no blurring and ghosting artifacts and good contrast between the rootlets and cerebrospinal fluid); coverage from the pontomedullary junction (PMJ) to the T1 vertebra; and available information on the participant's height. Combining all three datasets, a total of 120 healthy participants were used.

Figure 2 illustrates the analysis pipeline for assessing spinal-vertebral level correspondence, performed using the SCT v7.0[19]. Spinal cord segmentations were available for all T2w MRI scans[20,25], whereas the MP2RAGE scans were segmented using the contrast-agnostic model[29,30]. Spinal levels were estimated as an overlap between the spinal rootlets and the spinal cord segmentation dilated by 3 pixels[14]. The spinal cord centerline was extracted from the spinal cord segmentation, and the posterior tips of intervertebral discs[10] were projected to the centerline. Vertebral levels were determined as the segments between adjacent intervertebral discs. PMJ was detected automatically[31] and used as a reference point for further measurements.

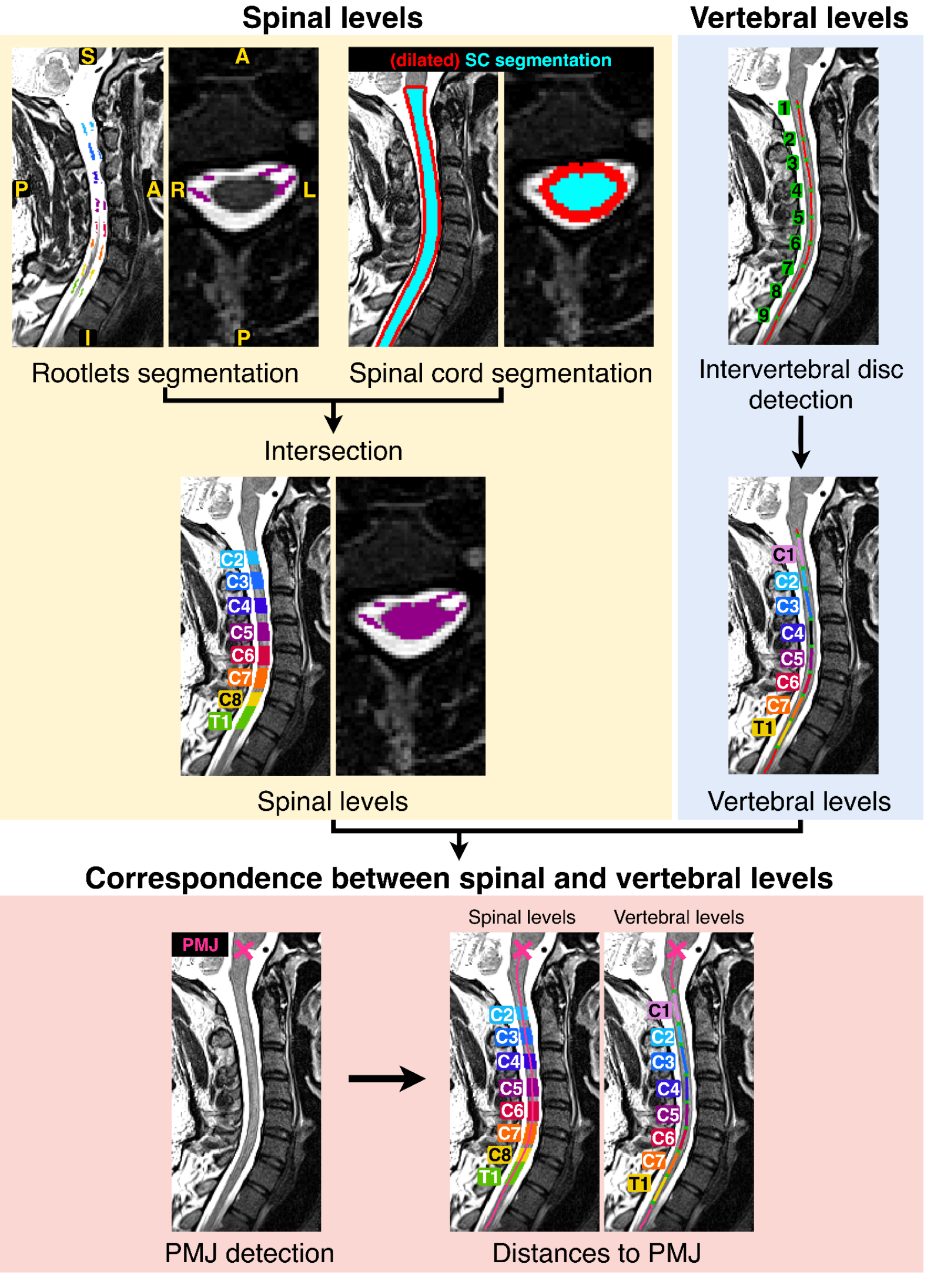

**Figure 2: Spinal-vertebral levels correspondence.** Analysis overview: the spinal cord and nerve rootlets were segmented, and the intervertebral discs and the pontomedullary junction (PMJ) were identified. Then, rootlets and intervertebral discs were used to obtain the spinal and vertebral levels, respectively. Distances from level midpoints to the PMJ were measured along the centerline.

The distances between the PMJ and the midpoints of the spinal and vertebral levels, as well as the level lengths, were measured along the spinal cord centerline. Spinal and vertebral level lengths were defined as the distance between the rostro-caudal slices of each level. To account for individual differences in body size, the distances were normalized by the participant's height and then multiplied by the cohort's median height to preserve millimetre units. The normalized distances were approximated by normal distributions using the probability density function to assess the overlap between spinal and vertebral levels. The correspondence between spinal and vertebral levels was evaluated using the Wilcoxon signed-rank test, the Bland-Altman analysis and root mean square distance (RMSD), separately for each level pair (e.g., vertebral level C2 and spinal level C3) across participants; Equation 1 and 2 in the Supplementary material (1.3. Spinal-vertebral Levels Correspondence and Level Lengths). We used a non-parametric variant of Bland-Altman analysis because the differences between spinal and vertebral level midpoint positions did not pass the Shapiro-Wilk normality test. $P < .05$ was considered statistically significant.

# Results

## Patient Characteristics

A total of 134 healthy adult participants with 177 MRI scans from three datasets were included in this study. Participants were scanned across scanners from different manufacturers (Siemens, GE, Philips) with different field strengths (3T, 7T). For segmentation model development, we used 50 participants with 93 MRI scans (n=12 OpenNeuro, n=24 SpineGeneric and n=57 MP2RAGE dataset) with 76 scans in the training set, and 17 scans in the testing set. For spinal-vertebral correspondence analysis, we used 120 MRI scans (n=4 OpenNeuro, n=105 SpineGeneric and n=11 MP2RAGE) from 120 participants; 14 participants were not included because their MRI scans did not capture the PMJ and/or the scans did not capture the whole T1 vertebra.

## Segmentation Model

The *RootletSeg* nnUNetv2 3D model achieved an overall Dice score of 0.65 ± 0.10 and HD95 of 5.10 ± 1.82 mm (mean ± standard deviation [SD] across levels, participants, and contrasts). Lower Dice for the MP2RAGE scans at levels C2 and C3 was due to the presence of image artifacts (red arrows in Figure 3), possibly caused by B1+ inhomogenities. Interestingly, despite the artifacts, the model was able to segment some rootlets at these levels, even though they were not included in the reference standard (compare the reference standard and model output for the T1w-INV2 MRI scan at the C2 level in Figure 3). Dice scores for individual contrasts were (mean ± SD across levels and testing scans): 0.67 ± 0.09 for T1w-INV2, 0.65 ± 0.11 for UNIT1, 0.64 ± 0.08 for T2w, and 0.62 ± 0.10 for T1w-INV1 (Figure 4a); and HD95 were: 4.70 ± 1.55 mm for T1w-INV2, 4.88 ± 1.60 mm for UNIT1, 6.20 ± 2.87 mm for T2w and 5.21± 1.75 mm for T1w-INV1 (Figure 4b). Figure 4c and Figure 4d show the Dice scores and HD95 values across levels on T2w MRI scans, in comparison with an existing model developed exclusively on T2w scans. *RootletSeg* demonstrated comparable results for rootlets C2-C8 to the T2w model (Dice of 0.65 ± 0.08 vs. 0.64 ± 0.08; HD95 of 5.65 ± 2.13 mm vs. 6.07 ± 3.75 mm). Level T1 was excluded from the Dice and HD95 calculation, as the T2w model was not trained at this level. The total inference runtime,  measured on a single sample T2w scan, was 13.67 s on an NVIDIA GeForce RTX 3080 GPU 124 GB RAM and 639.63 s on an Apple M3 16 GB RAM.

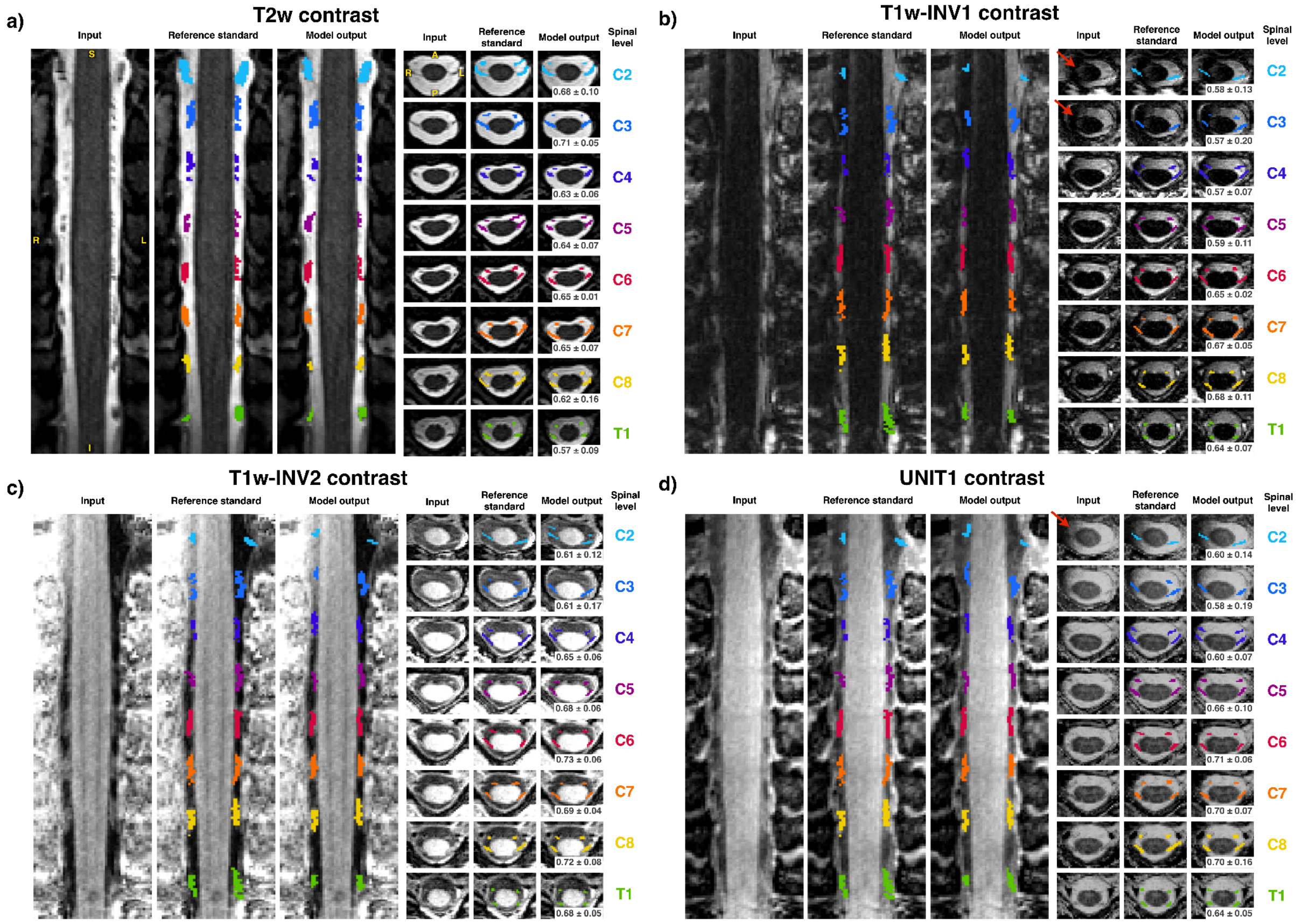

**Figure 3: Coronal and axial views of representative rootlet segmentations. (a)** Segmentations on a T2w MRI scan. **(b)** Segmentations on a T1w-INV1 MRI scan. **(c)** Segmentations on a T1w-INV2 MRI scan. **(d)** Segmentations on a UNIT1 MRI scan. The red arrows point to an artifact possibly caused by B1+ inhomogeneities. Rows represent individual rootlet levels from C2 to T1. Numbers represent the mean ± SD Dice score across participants for each spinal level compared to the reference standard labels.

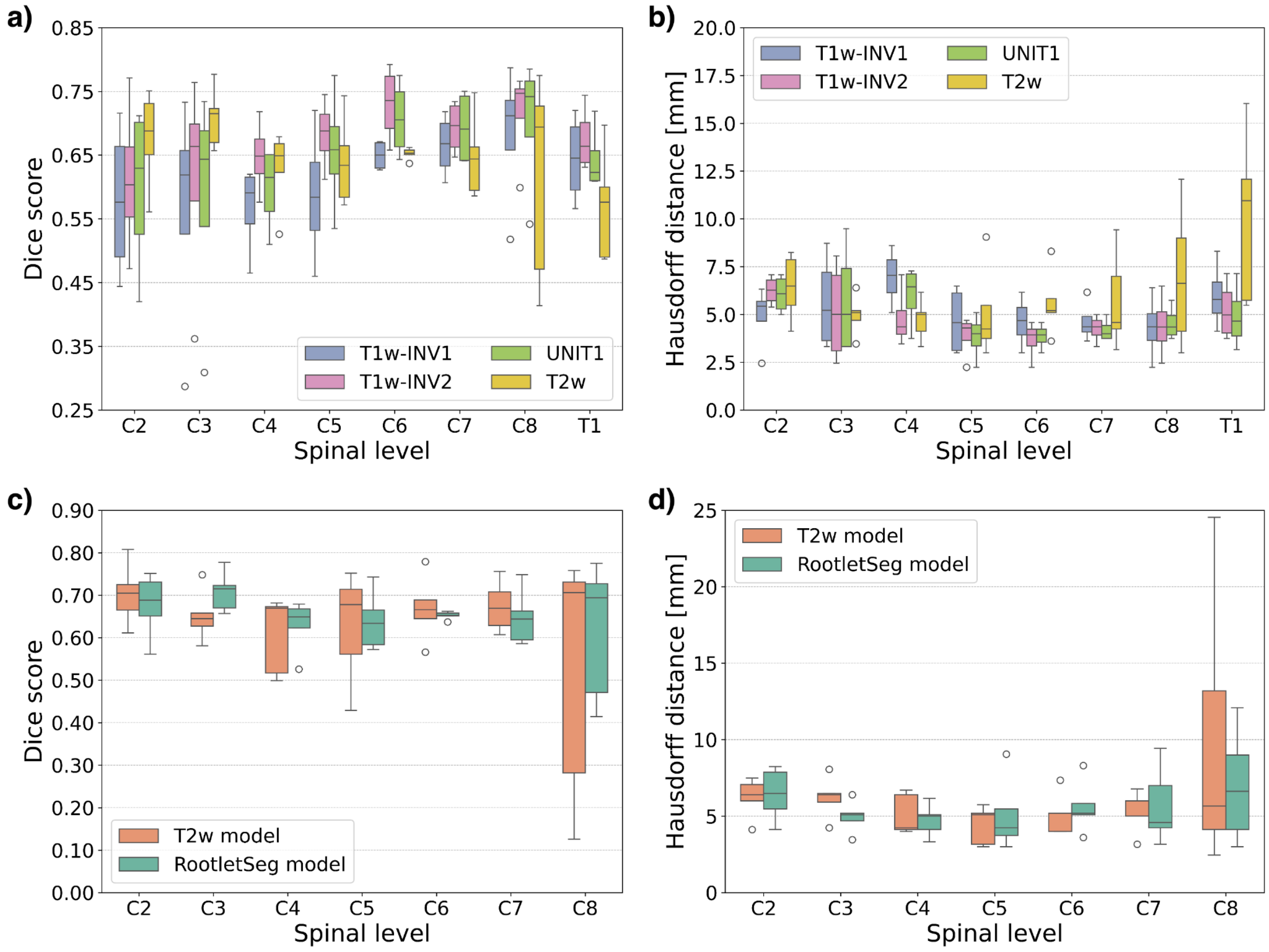

**Figure 4: Quantitative performance of the *RootletSeg* model. (a–b)** Dice score and HD95 computed on 17 testing images across different contrasts (n=4 T1w-INV1, n=4 T1w-INV2, n=4 UNIT1 and n=5 T2w). **(c–d)** Comparison of Dice score and HD95 between the intermediate single-contrast T2w model and the proposed *RootletSeg* model, on T2w MRI scans. Despite the *RootletSeg* model being trained on multiple contrasts, it performs similarly well compared to the T2w model. Note that the T2w model was developed only for spinal rootlets C2-C8; thus, T1 rootlets are not included in the comparison. The horizontal line within each box represents the median, and the box edges mark the 25 % to 75 % interquartile range. Whiskers extend 1.5 times the interquartile range, and the small black circles indicate outliers.

## Spinal-vertebral Level Analysis

Figure 5a illustrates the correspondence between spinal and vertebral levels. A gradually increasing *shift* along the rostrocaudal axis is apparent between the distributions of spinal and vertebral levels. For instance, the distribution for vertebral level C2 overlaps with that of spinal level C3, whereas vertebral level C7 is shifted caudally relative to spinal level C8. The Wilcoxon signed-rank test (performed for each spinal-vertebral level pair separately) revealed that this shift was statistically significant ($P < .05$) for all levels below the spinal level C5 and the vertebral level C4. Figure 5b presents the Bland-Altman analysis comparing each pair of levels (e.g., vertebral level C2 vs. spinal level C3), based on the distance of the level midpoints from the PMJ, normalized by participant height. The Bland-Altman analysis quantitatively assessed a 0.00 mm bias term (median difference between spinal and vertebral level midpoint positions) between spinal level C3 and vertebral level C2. In contrast, the bias term is higher for the lower levels (up to 8.15 mm for spinal level T1 and vertebral level T1). Suppl. Table 2 shows the RMSD between spinal and vertebral level midpoints for each level pair. The RMSD value was lowest for spinal level C2 and vertebral level C3 (3.60 mm) and highest for spinal level T1 and vertebral level T1 (9.36 mm).

Suppl. Table 3 and Suppl. Table 4 show the rostro-caudal lengths of each spinal and vertebral level (mean ± SD across participants) in 120 healthy participants. Suppl. Table 3 includes a comparison with lengths reported in an MRI-based study[8] and a post-mortem study[32], and Suppl. Table 4 with a comparison to a post-mortem study [33].

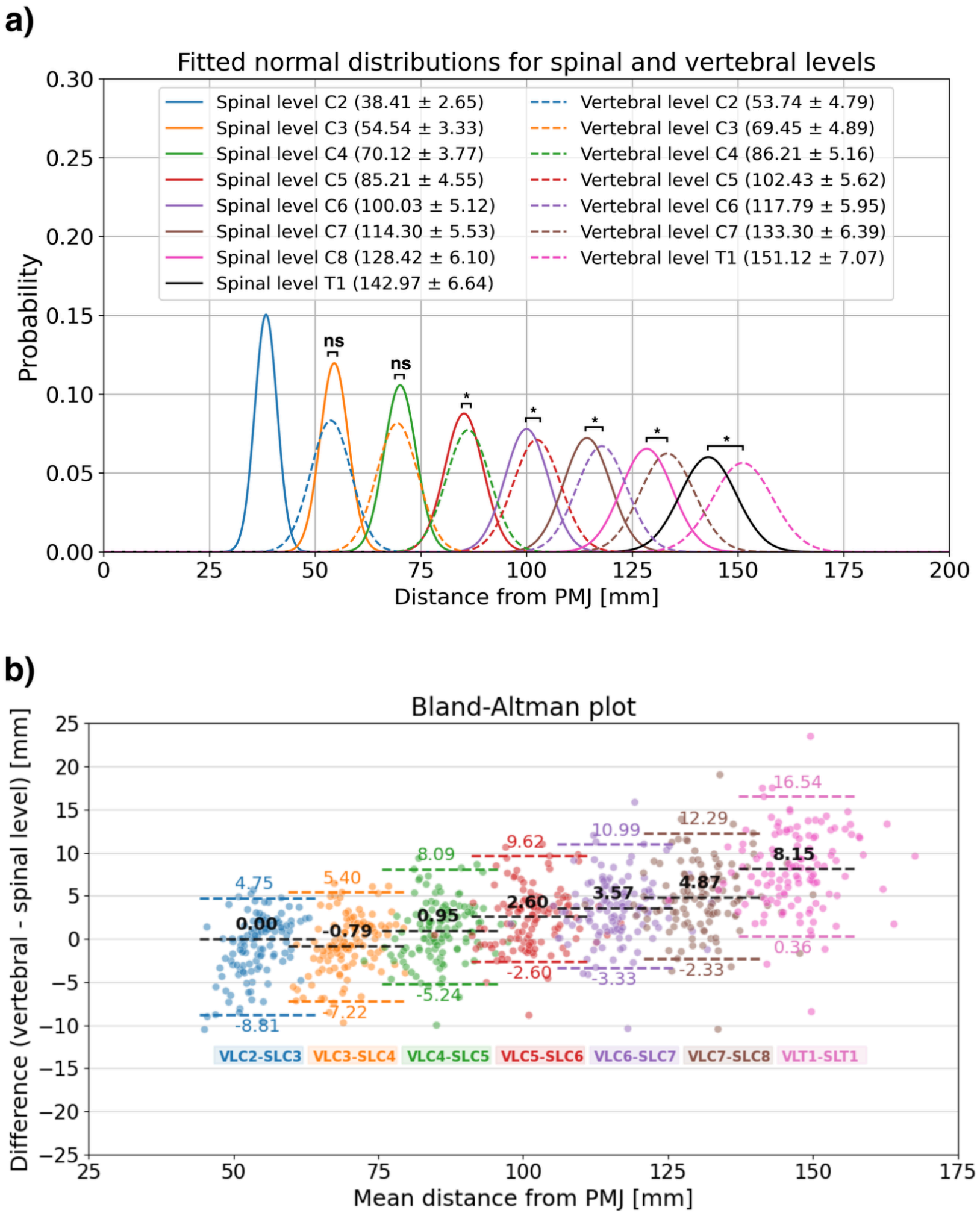

**Figure 5: Spinal and vertebral level correspondence. (a)** Spinal and vertebral level midpoints approximated by normal distributions, separately for each level. The midpoints were normalized by participants' height and scaled by median height. Values in brackets represent mean ± SD distance to the pontomedullary junction (PMJ) in millimetres. Spinal levels are in solid lines, vertebral levels in dashed lines. Significance (Wilcoxon signed-rank test): * $P$ < .05, *ns* not significant. Notice that the distribution for the spinal level C3 (solid orange) corresponds to the vertebral level C2 (dashed blue), while the distribution for the spinal level C8 (solid pink) is shifted cranially relative to the vertebral level C7 (dashed brown). We note that there are anatomically seven vertebral levels but eight spinal levels. **(b)** Bland-Altman plot. Black dashed lines show the median difference between distances from the PMJ to spinal and vertebral levels midpoints, and colored dashed lines show 2.5 and 97.5 percentiles. The points correspond to individual participants. VL = vertebral level; SL = spinal level.

# Discussion

In this study, we (i) introduced *RootletSeg*, a deep learning model for segmentation of C2-T1 ventral and dorsal spinal nerve rootlets from T1w, T2w and MP2RAGE-UNIT1 3T and 7T MRI scans, which extended the previous rootlet segmentation model[14] by incorporating ventral rootlets, an additional thoracic T1 spinal level, and additional 7T MP2RAGE-derived contrasts; and (ii) investigate the correspondence between spinal and vertebral levels in a large cohort of 120 healthy participants. The segmentation model demonstrated stable performance across participants, MRI contrasts, and rootlet levels, thus facilitating the cumbersome and time-intensive manual rootlets annotation process. The analysis of spinal-vertebral level correspondence showed a gradually increasing shift in the rostro-caudal axis between spinal and vertebral levels and higher variability in level localization across participants with lower levels. The segmented nerve rootlets can be used as an alternative to commonly used intervertebral discs in various applications, including lesion classification based on neurological levels and registration of individual scans to a template for group-level analysis[5,6].

As spinal nerve rootlets are fine structures with submillimeter dimensions and a complex anatomy that varies across participants, their segmentation is challenging even for expert raters. Despite these difficulties, the proposed model achieved a stable performance across four contrasts (T2w, T1w-INV1, T1w-INV2, and UNIT1) and different levels (C2-T1). For T2w data, the mean Dice was higher and HD95 was lower for upper cervical rootlets relative to lower rootlets (C2 mean Dice: 0.68 ± 0.08 vs. T1 mean Dice: 0.57 ± 0.09; C2 mean HD95: 6.44 ± 1.70 vs. T1 mean HD95: 6.97 ± 3.68), possibly due to the lower contrast between cerebrospinal fluid and rootlets and higher rootlets angulation in the lower levels[13]. Compared to the intermediate single-contrast T2w model, *RootletSeg* achieved a comparable Dice of 0.65 ± 0.08 vs. 0.64 ± 0.08 and a better HD95 of 5.65 ± 2.13 mm vs. 6.07 ± 3.75 mm, demonstrating no loss in performance while extending the model capabilities beyond T2w contrast. Although the reported segmentation performance may appear lower compared to other segmentation tasks, such as spinal cord[30,34] and spinal canal[35], where Dice scores commonly reach 0.9, we note that the relatively low Dice and high HD95 for rootlet segmentation is due to the distinct anatomy and size of rootlets compared to larger structures like the spinal cord. Spinal rootlets are small structures with complex three-dimensional anatomy, typically having only 2–3 voxel width in MRI scans with submillimeter in-plane resolution.

The analysis of spinal and vertebral level correspondence using the *RootletSeg* model showed that the distribution of spinal level C3 midpoint positions corresponds to that of vertebral level C2, and similarly for spinal level C4 and vertebral level C3. The correspondence became less consistent at lower levels, as indicated by a statistically significant shift between spinal and vertebral levels, leading to larger shifts between spinal and vertebral midpoint positions. Moreover, SD of the level midpoint distances from the PMJ increases in the lower levels (spinal level C2 SD = 2.65 mm vs. spinal level T1 SD = 6.64 mm), resulting in broader and flatter distributions, indicating increasing inter-subject variability in level positions. This is consistent with prior MRI and post-mortem reports[8,9] and anatomical textbooks that neurological spinal levels are "*shifted*" relative to vertebral levels and that this shift increases at more caudal levels. Similar to a previous 3T MRI study[8] that used manually defined landmarks, the Bland-Altman analysis showed higher variability in the position of lower levels across the participants. In our study, the analysis was extended to include levels C2 and T1, and we used 6 times more data. Additionally, we used participants' height to normalize our measurements to take into account inter-subject variability. We also analyzed the level correspondence using RMSD, which confirmed a higher shift for more caudal levels by higher RMSD values compared to more cranial levels. The difference between the Bland-Altman and the RMSD analyses (e.g., Bland-Altman bias of 0.00 mm and RMSD of 3.60 mm for vertebral level C2 and spinal level C3) is due to methodological differences in the calculation − in the Bland-Altman analysis, we quantitatively considered the correspondence according to the median difference between vertebral-spinal midpoint positions, whereas the RMSD was calculated as the mean squared difference. A post-mortem study[9], performed manually on 16 cadavers, found that spina-vertebral correspondence differs by one level in the C3 to C5 spinal region, i.e., vertebral level C2 corresponds to spinal level C3 and further increases in the lower levels up to two level differences. It needs to be noted that it is difficult to directly compare the findings from post-mortem studies with those in our in vivo MRI study due to the inherent characteristics of ex vivo measures, such as tissue shrinking due to post-fixation, and the altered shape of the excised spinal cord without the surrounding cerebrospinal fluid and dura.

Measured rostrocaudal spinal level lengths obtained in our study showed slightly higher SD compared to an MRI-based study[8] and a post-mortem study[32]. This might likely be due to the larger cohort in our study capturing broader population variability, demographic differences, and differences in the acquisition protocol. Due to the lack of MRI studies on vertebral level lengths, our results were compared with a post-mortem study[33], which measured vertebral levels from CT scans in six cadavers. For levels C3 to T1, our findings show similar results to the study. However, they measured the length of vertebral bodies, whereas our analysis was based on intervertebral disc positions. This different methodological approach likely contributes to the average difference of 2.1 mm observed across levels. Additionally, the small sample size (six cadavers in the study) may not adequately capture the population variability. Other factors that may account for the demographic differences and positional changes of the spine structures between living participants and cadavers.

This study had limitations. The proposed model was trained and tested solely on healthy participants in the C2-T1 region and only with isotropic MRI scans. Scans were selected based on overall image quality, and scans with extensive blurring and ghosting artifacts were excluded to allow for reliable reference standard labels creation. Extending the evaluation to participants with pathologies, including spinal cord injury or multiple sclerosis, and the thoraco-lumbar region would be valuable. However, this remains challenging due to the lower contrast typically present in pathological areas, such as a narrowed spinal canal. Additionally, in the thoraco-lumbar spinal cord, rootlets are more difficult to isolate because of their steeper angulation, reduced space within the spinal canal, potential image artifacts due to respiratory motion, and lower signal-to-noise ratio at 7T. These factors make it difficult to obtain reliable reference standard labels. For the spinal-vertebral correspondence analysis, we used the participants' height to account for potential biological differences among individuals. The spinal cord or spine length could also be considered[11], but MRI data typically does not cover the entire spine. We performed the level correspondence analysis on a population of healthy adults in the cervical region only, without distinguishing differences between males and females and between different age groups. Spinal–vertebral correspondence analysis relied on spinal cord and rootlet segmentations and may thus be influenced by segmentation accuracy. In the caudal region, the angulation of rootlets changes and a single axial image slice can contain rootlets from multiple spinal levels, making it difficult to reliably infer spinal levels using the intersection of rootlet and spinal cord segmentations. To address this increased complexity in these regions, it would be advantageous to propose a more robust method for obtaining spinal levels from rootlets segmentation[6].

# Conclusion

In conclusion, this study presented *RootletSeg*, a deep learning model for C2-T1 spinal rootlets segmentation on T1w, T2w and MP2RAGE-UNIT1 MRI scans. The segmentation method is open-source, implemented in the `sct_deepseg` function as part of Spinal Cord Toolbox v7.0 and higher[19]. As the *RootletSeg* model allows for inferring the spinal levels directly from MRI, it can facilitate various downstream analyses, including lesion classification, neuromodulation therapy, and fMRI group analysis.

# References

1. Mohajeri Moghaddam S, Bhatt AA. Location, length, and enhancement: systematic approach to differentiating intramedullary spinal cord lesions. *Insights Imaging* 2018;9:511–26.

2. Ahuja CS, Wilson JR, Nori S, et al. Traumatic spinal cord injury. *Nat Rev Dis Primers* 2017;3.

3. Vallejo R MD, PhD. Neuromodulation of the cervical spinal cord in the treatment of chronic intractable neck and upper extremity pain: A case series and review of the literature. *Pain Physician* 2007;2;10:305–11.

4. Rowald A, Komi S, Demesmaeker R, et al. Activity-dependent spinal cord neuromodulation rapidly restores trunk and leg motor functions after complete paralysis. *Nat Med* 2022;28:260–71.

5. Kinany N, Pirondini E, Micera S, et al. Spinal Cord fMRI: A New Window into the Central Nervous System. *Neuroscientist* 2023;29:715–31.

6. Bédard S, Valošek J, Oliva V, et al. Rootlets-based registration to the spinal cord PAM50 template. *Imaging Neuroscience* https://doi.org/10.1162/IMAG.a.123.

7. Schlienger R, Landelle C, Hernandez-Charpak SD, et al. Mapping Human Proprioceptive Projections of Upper Limb Muscles Through Spinal Cord fMRI. *Hum Brain Mapp* 2025;46:e70386.

8. Cadotte DW, Cadotte A, Cohen-Adad J, et al. Characterizing the location of spinal and vertebral levels in the human cervical spinal cord. *AJNR Am J Neuroradiol* 2015;36:803–10.

9. Kim JH, Lee CW, Chun KS, et al. Morphometric Relationship between the Cervicothoracic Cord Segments and Vertebral Bodies. *J Korean Neurosurg Soc* 2012;52:384–90.

10. Ullmann E, Pelletier Paquette JF, Thong WE, et al. Automatic labeling of vertebral levels using a robust template-based approach. *Int J Biomed Imaging* 2014;2014:719520.

11. Frostell A, Hakim R, Thelin EP, et al. A Review of the Segmental Diameter of the Healthy Human Spinal Cord. *Front Neurol* 2016;7:238.

12. Diaz E, Morales H. Spinal Cord Anatomy and Clinical Syndromes. *Semin Ultrasound CT MR* 2016;37:360–71.

13. Mendez A, Islam R, Latypov T, et al. Segment-Specific Orientation of the Dorsal and Ventral Roots for Precise Therapeutic Targeting of Human Spinal Cord. *Mayo Clin Proc* 2021;96:1426–37.

14. Valošek J, Mathieu T, Schlienger R, et al. Automatic segmentation of the spinal

cord nerve rootlets. *Imaging Neurosci (Camb)* 2024;2:1–14.

15. Dauleac C, Frindel C, Pélissou-Guyotat I, et al. Full cervical cord tractography: A new method for clinical use. *Front Neuroanat* 2022;16:993464.

16. Liu J, Zhang W, Zhou Y, et al. An open-access lumbosacral spine MRI dataset with enhanced spinal nerve root structure resolution. *Sci Data* 2024;11:1131.

17. Fasse A, Newton T, Liang L, et al. A novel CNN-based image segmentation pipeline for individualized feline spinal cord stimulation modeling. *J Neural Eng* 2024;21:036032.

18. Liang L, Fasse A, Damiani A, et al. SpIC3D imaging: Spinal In-situ Contrast 3D imaging. *bioRxiv* https://doi.org/10.1101/2025.03.05.641747.

19. De Leener B, Lévy S, Dupont SM, et al. SCT: Spinal Cord Toolbox, an open-source software for processing spinal cord MRI data. *Neuroimage* 2017;145:24–43.

20. Bédard S, Bouthillier M, Cohen-Adad J. Pontomedullary junction as a reference for spinal cord cross-sectional area: validation across neck positions. *Sci Rep* 2023;13:13527.

21. Cohen-Adad J, Alonso-Ortiz E, Abramovic M, et al. Author Correction: Open-access quantitative MRI data of the spinal cord and reproducibility across participants, sites and manufacturers. *Sci Data* 2021;8:251.

22. Horn U, Vannesjo SJ, Gross-Weege N, et al. Ultra-high-field fMRI reveals layer-specific responses in the human spinal cord. *bioRxiv* https://doi.org/10.1101/2025.07.17.665316.

23. Massire A, Taso M, Besson P, et al. High-resolution multi-parametric quantitative magnetic resonance imaging of the human cervical spinal cord at 7T. *Neuroimage* 2016;143:58–69.

24. Cohen-Adad J, Alonso-Ortiz E, Abramovic M, et al. Generic acquisition protocol for quantitative MRI of the spinal cord. *Nat Protoc* 2021;16:4611–32.

25. Cohen-Adad J, Alonso-Ortiz E, Abramovic M, et al. Open-access quantitative MRI data of the spinal cord and reproducibility across participants, sites and manufacturers. *Sci Data* 2021;8:219.

26. Isensee F, Jaeger PF, Kohl SAA, et al. nnU-Net: a self-configuring method for deep learning-based biomedical image segmentation. *Nat Methods* 2021;18:203–11.

27. Tubbs RS, Loukas M, Slappey JB, et al. Clinical anatomy of the C1 dorsal root, ganglion, and ramus: a review and anatomical study. *Clin Anat* 2007;20:624–7.

28. Maier-Hein L, Reinke A, Godau P, et al. Metrics reloaded: recommendations for image analysis validation. *Nat Methods* 2024;21:195–212.

29. Bédard S, Karthik EN, Tsagkas C, et al. Towards contrast-agnostic soft segmentation of the spinal cord. *Med Image Anal* 2025;101:103473.

30. Karthik EN, Sandrine B, Jan V, et al. Monitoring morphometric drift in lifelong learning segmentation of the spinal cord. *arXiv [csCV]* 2025 May 2. [Epub ahead of print].

31. Gros C, De Leener B, Dupont SM, et al. Automatic spinal cord localization, robust to MRI contrasts using global curve optimization. *Med Image Anal* 2018;44:215–27.

32. Kobayashi R, Iizuka H, Nishinome M, et al. A cadaveric study of the cervical nerve roots and spinal segments. *Eur Spine J* 2015;24:2828–31.

33. Busscher I, Ploegmakers JJW, Verkerke GJ, et al. Comparative anatomical dimensions of the complete human and porcine spine. *Eur Spine J* 2010;19:1104–14.

34. Gros C, De Leener B, Badji A, et al. Automatic segmentation of the spinal cord and intramedullary multiple sclerosis lesions with convolutional neural networks. *Neuroimage* 2019;184:901–15.

35. Salmona, A. Bédard, S., Valosek J., et al. Automated robust segmentation of the spinal canal on MRI. European Journal of Radiology Artificial Intelligence 6, 100075 (2026).

# Acknowledgments

We thank Mathieu Guay-Paquet, Joshua Newton, and Kalum Ost for their assistance with the management of the datasets and the implementation of the algorithm in the Spinal Cord Toolbox. We thank Nathan Molinier for the help with organizing the dataset according to the BIDS standard. We thank Louis-Thomas Lapointe for the help with the data annotation and preliminary model training.

# Data and Code Availability

The analysis scripts are open source and available at: https://github.com/ivadomed/model-spinal-rootlets/releases/tag/r20250917. The packaged and ready-to-use *RootletSeg* model can be applied to custom data via the `sct_deepseg rootlets -i <MRI-scan>` command as part of the Spinal Cord Toolbox (SCT) v7.0 and higher: https://github.com/spinalcordtoolbox/spinalcordtoolbox/releases/tag/7.0.

The data come from open-access datasets and can be accessed at https://openneuro.org/datasets/ds004507/versions/1.1.1 and https://github.com/spine-generic/data-multi-subject/tree/r20250314. The data from the private MP2RAGE dataset will be shared upon reasonable request.

**Funding information**

Funded by the Canada Research Chair in Quantitative Magnetic Resonance Imaging [CRC-2020-00179], the Canadian Institute of Health Research [PJT-190258, PJT-203803], the Canada Foundation for Innovation [32454, 34824], the Fonds de Recherche du Québec - Santé [322736, 324636], the Natural Sciences and Engineering Research Council of Canada [RGPIN-2019-07244], the Canada First Research Excellence Fund (IVADO and TransMedTech), the Courtois NeuroMod project, the Quebec BioImaging Network [5886, 35450], INSPIRED (Spinal Research, UK; Wings for Life, Austria; Craig H. Neilsen Foundation, USA), Mila - Tech Transfer Funding Program. JV received funding from the European Union's Horizon Europe research and innovation programme under the Marie Skłodowska-Curie grant agreement No 101107932 and is supported by the Ministry of Health of the Czech Republic (no. NU22-04-00024). SB is supported by the Natural Sciences and Engineering Research Council of Canada, NSERC, Canada Graduate Scholarships — Doctoral program. Computational resources were provided by the e-INFRA CZ project (ID:90254), supported by the Ministry of Education, Youth and Sports of the Czech Republic. FE received funding from the European Research Council (under the European Union's Horizon 2020 research and innovation programme; grant agreement No 758974) and the Max Planck Society.

**Supplementary material**

**Title**

**Segmentation of spinal rootlets across MRI contrasts with RootletSeg**

# Materials And Methods

## Study Design and Participants

The scans were anonymized and defaced to remove all personally identifiable information (1,2) and organized according to the BIDS standard (3). The inclusion criteria were being a healthy participant without any known disease, age >18 years, and coverage of the cervical spine. Exclusion criteria were the presence of severe imaging artifacts or low contrast between the cerebrospinal fluid and nerve rootlets for multiple levels. Although the original datasets include more participants than listed in Suppl. Table 1, we selected scans based on overall scan quality (no blurring and ghosting artifacts) and sufficient contrast between the cerebrospinal fluid and nerve rootlets. Most exclusions were due to lower image quality, caused mainly by reduced spinal canal space or artifacts likely originating from B1+ inhomogenities (8 participants from the MP2RAGE dataset). Additional scans were excluded because the T1 spinal level was not included in the field of view (FOV) (11 participants), the pontomedullary junction (PMJ) was not in the FOV (7 participants), or height information was missing (2 participants). Because manual segmentation of nerve rootlets is difficult and time-consuming due to their complex three-dimensional anatomy (Suppl. Figure 1) and the scans' sub-millimetre resolution, only a subset of participants was included in model training (see the following sections for details).

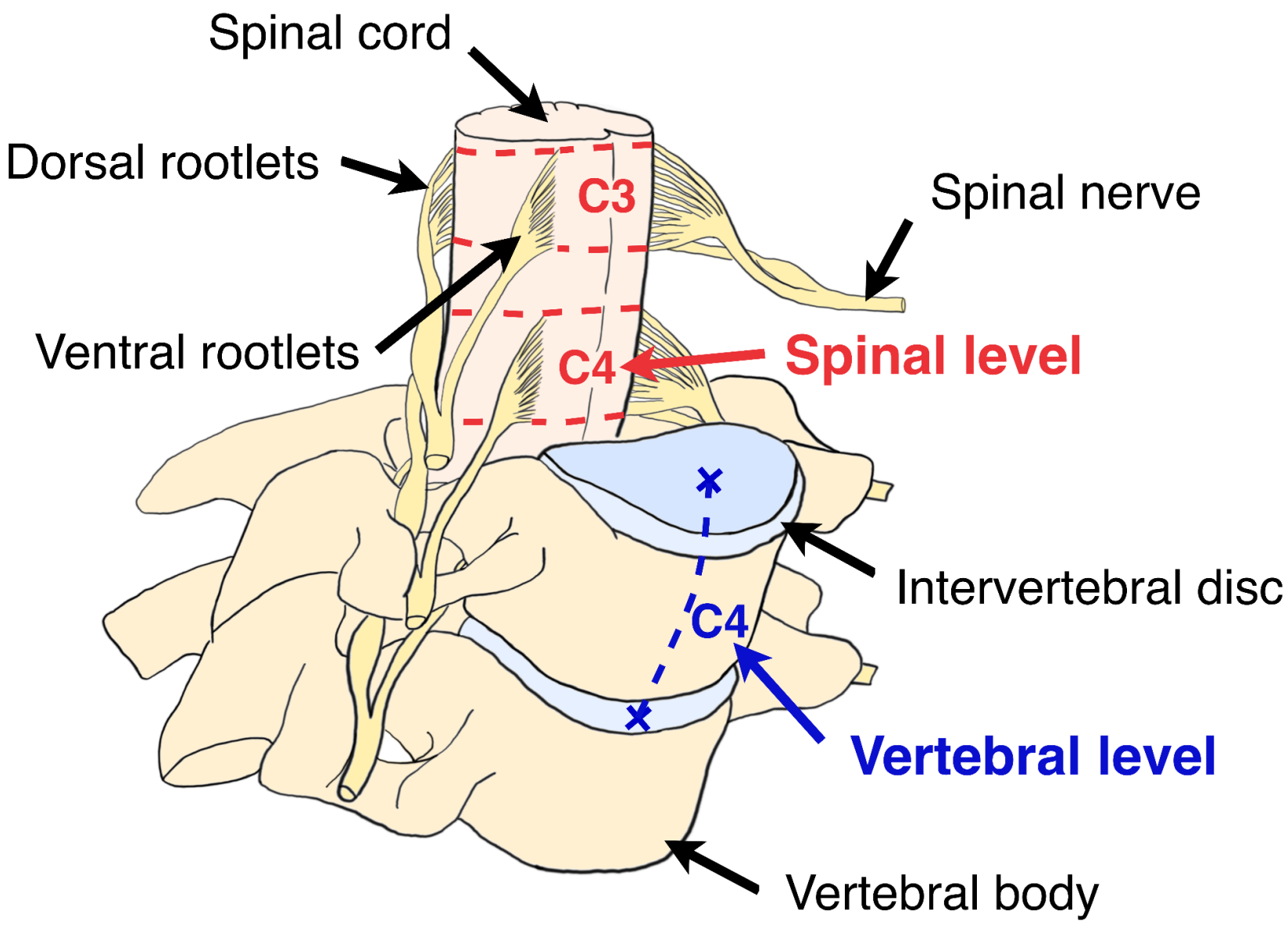

**Supplementary Figure 1: Spinal rootlets and vertebral anatomy.** Spinal levels are inferred from spinal rootlets, whereas vertebral levels are defined based on adjacent intervertebral discs. Adapted from (4) with permission from the publisher.

## Deep Learning Training Protocol

Images and corresponding reference standard labels were reoriented to the Right-to-Left × Posterior-to-Anterior × Inferior-to-Superior (RPI) orientation to ensure consistency across the datasets. Preprocessing with the nnUNetv2 framework involved intensity z-score normalization and resampling into 0.7 × 0.7 × 0.7 mm resolution before model training. Random spatial transformations (translation, rotation, scaling), mirroring, Gaussian noise and Gaussian blur, brightness and contrast adjustment, Gamma transformation and low-resolution simulation were used for data augmentation.

We extended the T2w model to segment ventral rootlets by manually annotating them and retraining the model. Then, the T2w model was applied to MP2RAGE-UNIT1 scans. We inverted the contrast of MP2RAGE-UNIT1 scans to make their contrast closer to the T2w before applying the model. The model predictions were manually corrected and extended to include T1 rootlets. An initial MP2RAGE model was trained using five scans (MP2RAGE-UNIT1 with inverted contrast) and tested on the remaining 14 scans from the MP2RAGE dataset. The obtained segmentations were manually corrected to get a reference standard for all 19 participants. Once all rootlet reference standards were visually inspected and corrected, they were split into training, validation, and testing sets. Because the MP2RAGE contrasts (T1w-INV1, T1w-INV2, and UNIT1) are inherently co-registered, a single reference standard was used for all three contrasts for each participant. To prevent information leakage between training, validation and testing sets, all MP2RAGE contrasts from each participant were assigned exclusively to one set.

Then, we trained the model on all three MP2RAGE contrasts (15 participants, resulting in 45 MRI scans) for 2000 epochs. The nnUNetv2 framework suggested a 3D architecture with the following parameters: 6 stages, instance normalization technique, Leaky ReLU activation function, batch size 2, patch size 128 × 96 × 192 (RPI), learning rate 0.01, Dice Loss and Cross-Entropy loss function and Stochastic Gradient Descent with Nesterov Momentum optimizer. Since training with this architecture setup was not successful at the C2 level, we tried another experiment with increased patch size in the superior-inferior direction to 352 (i.e., 128 × 96 × 352) so the model can capture a larger spatial context. With this increased patch size, the model successfully learned to segment the C2 level. The model training was consequently extended to a multi-contrast approach, adding data from T2w datasets (spine-generic and OpenNeuro). This model was trained on MP2RAGE and T2w MRI scans with the increased patch size (i.e., 128 × 96 × 352). A total of 76 scans (from 50 participants) for 2000 epochs in a 5-fold cross-validation approach were used for an 80/20% training/validation split. The “production” model, named RootletSeg, was trained on all 76 training data (100/0% training/validation split). The testing set (i.e., scans never used during the training or validation) included 17 MRI scans: 12 MP2RAGE (4 participants, each with 3 contrasts) and 5 T2w (3 from the SpineGeneric dataset and 2 from OpenNeuro). The same testing set was used across all five cross-validation folds to enable consistent evaluation of model performance across varied training/validation splits. The train–test split was performed at the participant level. Consequently, no participant overlap occurred between the training and testing sets across contrasts. The model was compared to the intermediate T2w single-contrast model.

# Spinal-vertebral Levels Correspondence and Level Lengths

$$BlandAltman\ median\ =\ median(y_{vertebral\ (1)} - y_{spinal(1)},\ y_{vertebral\ (i+1)} - y_{spinal(i+1)},\ \dots\ ,\ y_{vertebral\ (n)} - y_{spinal(n)})$$

**Equation 1**

$y_{spinal(i)}$: distance between the PMJ and the spinal level midpoint

$y_{vertebral(i)}$: distance between the PMJ and the vertebral level midpoint

n … number of participants

$$RMSD = \sqrt{\sum_{i=1}^{n} \frac{(y_{vertebral(i)} - y_{spinal(i)})^2}{n}}$$

**Equation 2**

$y_{spinal(i)}$: distance between the PMJ and the spinal level midpoint

$y_{vertebral(i)}$: distance between the PMJ and the vertebral level midpoint

n: number of participants

# Tables

| **RootletSeg model development** | | | |
|---|---|---|---|
| **Variable** | **OpenNeuro ds004507*** | **spine-generic multi-subject** | **MP2RAGE**** |
| **Participants** | 7 | 24 | 19 |
| **MRI scans** | 12 | 24 | 57 |
| **Sex** | | | |
| Male | 5 | 12 | 11 |
| Female | 2 | 12 | 8 |
| Age (y) | (22.57 ± 0.53) | (29.58 ± 6.53) | (29.84 ± 6.66) |
| **MRI scans in each set** | | | |
| Training set | 10 | 21 | 45 |
| Test set | 2 | 3 | 12 |
| **MRI manufacturer** | | | |
| Siemens | 12 | 20 | 57 |
| GE | 0 | 4 | 0 |
| **MRI field strength** | | | |
| 3T | 12 | 24 | 0 |
| 7T | 0 | 0 | 57 |
| **Sequence** | TSE | TSE | MP2RAGE |
| **Voxel size (mm)** | 0.6 × 0.6 × 0.6 | 0.8 × 0.8 × 0.8*** | 0.7 × 0.7 × 0.7 |

| **Spinal-vertebral level correspondence analysis** | | | |
|---|---|---|---|
| **Variable** | **OpenNeuro ds004507*** | **spine-generic multi-subject** | **MP2RAGE**** |
| Participants | 4 | 105 | 11 |
| MRI scans | 4 | 105 | 11 |
| **Sex** | | | |
| Male | 3 | 52 | 4 |
| Female | 1 | 53 | 7 |
| Age (y) | (22.50 ± 0.58) | (29.70 ± 10.18) | (28.82 ± 6.03) |
| **MRI manufacturer** | | | |
| Siemens | 4 | 76 | 11 |
| GE | 0 | 11 | 0 |
| Philips | 0 | 18 | 0 |
| **MRI field strength** | | | |
| 3T | 4 | 105 | 0 |
| 7T | 0 | 0 | 11 |
| **Sequence** | TSE | TSE | MP2RAGE |
| **Voxel size (mm)** | 0.6 × 0.6 × 0.6 | 0.8 × 0.8 × 0.8*** | 0.7 × 0.7 × 0.7 |

*The OpenNeuro ds004507 dataset included neutral, flexion and extension neck position sessions. Neutral and flexion sessions were used for the model development, and neutral session was used for spinal-vertebral correspondence analysis.
**The MP2RAGE dataset contained 3 co-registered MP2RAGE contrasts (T1w-INV1, T1w-INV2 and UNIT1) for each subject (in total 57 MRI scans for 19 subjects).
***Voxel size for 2 MRI scans was 0.8 × 0.5 × 0.5 mm

**Supplementary Table 1:** Characteristics of Study Participants

| Vertebral \| Spinal level | RMSD [mm] |
| --- | --- |
| C2 \| C3 | 3.60 |
| C3 \| C4 | 3.35 |
| C4 \| C5 | 3.55 |
| C5 \| C6 | 4.21 |
| C6 \| C7 | 5.09 |
| C7 \| C8 | 6.33 |
| T1 \| T1 | 9.36 |

**Supplementary Table 2:** Root mean square distance (RMSD) between spinal and vertebral level midpoint distances to the pontomedullary junction (PMJ). Notice that RMSD is lower for cranial levels (i.e., RMSD of 3.60 mm between the vertebral level C2 and the spinal level C3) relative to caudal levels (i.e., RMSD of 9.36 mm between the vertebral level T1 and the spinal level T1). We note that there are anatomically seven vertebral levels but eight spinal levels.

| **Spinal level** | **This work** (120 participants) | **Cadotte et al., 2015** (20 participants) | **Kobayashi et al., 2015** (11 participants) |
|---|---|---|---|
| C2 | 7.1 ± 2.2 | - | - |
| C3 | 11.7 ± 2.5 | 10.5 ± 2.2 | 12.1 ± 1.2 |
| C4 | 8.9 ± 2.0 | 9.9 ± 1.3 | 12.5 ± 1.1 |
| C5 | 8.7 ± 1.7 | 10.5 ± 1.5 | 12.6 ± 2.8 |
| C6 | 9.1 ± 1.6 | 9.7 ± 1.6 | 12.7 ± 1.6 |
| C7 | 9.8 ± 1.7 | 9.4 ± 1.4 | 11.8 ± 1.6 |
| C8 | 11.8 ± 2.5 | 9.6 ± 1.4 | 10.6 ± 1.6 |
| T1 | 13.1 ± 3.7 | - | - |

**Supplementary Table 3:** Rostro-caudal lengths of individual spinal levels and results of MRI-based study (4) and post-mortem study (5). The table shows the mean rostro-caudal length ± SD in millimetres.

| **Spinal level** | **This work** (120 participants) | **Busscher et al., 2010** (6 participants) |
|---|---|---|
| C2 | 14.9 ± 1.9 | - |
| C3 | 17.0 ± 1.8 | 14.2 ± 0.7 |
| C4 | 17.0 ± 1.6 | 14.5 ± 1.3 |
| C5 | 15.9 ± 1.5 | 13.4 ± 1.1 |
| C6 | 15.3 ± 1.7 | 14.0 ± 0.5 |
| C7 | 16.2 ± 1.7 | 15.7 ± 0.9 |
| T1 | 20.0 ± 1.8 | 17.3 ± 0.8 |

**Supplementary Table 4:** Rostro-caudal lengths of individual vertebral levels and results of the post-mortem study (6) (showing the mean rostro-caudal length ± SD in millimetres)

# Supplementary References

1. Li X, Morgan PS, Ashburner J, Smith J, Rorden C. The first step for neuroimaging data analysis: DICOM to NIfTI conversion. J Neurosci Methods. Elsevier; 2016;264:47–56.

2. Gulban OF, Nielson D, Poldrack R, et al. poldracklab/pydeface: v2.0.0. Zenodo; 2019. doi: 10.5281/ZENODO.3524401.

3. Gorgolewski KJ, Auer T, Calhoun VD, et al. The brain imaging data structure, a format for organizing and describing outputs of neuroimaging experiments. Sci Data. 2016;3:160044.

4. Cadotte DW, Cadotte A, Cohen-Adad J, et al. Characterizing the location of spinal and vertebral levels in the human cervical spinal cord. AJNR Am J Neuroradiol. 2015;36(4):803–810.

5. Kobayashi R, Iizuka H, Nishinome M, Iizuka Y, Yorifuji H, Takagishi K. A cadaveric study of the cervical nerve roots and spinal segments. Eur Spine J. 2015;24(12):2828–2831.

6. Busscher I, Ploegmakers JJW, Verkerke GJ, Veldhuizen AG. Comparative anatomical dimensions of the complete human and porcine spine. Eur Spine J. 2010;19(7):1104–1114.